\documentclass[aps,prl,10pt,twocolumn,superscriptaddress,nofootinbib,floatfix]{revtex4-2}
\usepackage{amsmath,amssymb,bm}
\usepackage{graphicx}
\usepackage{booktabs,tabularx,array}
\usepackage{xcolor}
\usepackage{microtype}
\usepackage[colorlinks=true,linkcolor=blue!50!black,citecolor=blue!50!black,urlcolor=blue!50!black]{hyperref}
\newcommand{\M}{\mathcal M}
\newcommand{\dd}{\mathrm d}
\newcommand{\e}{e}
\newcommand{\qT}{q_{\rm T}}
\newcommand{\qp}{q_{+}}
\newcommand{\ba}{b}

\begin{document}

%%%%%%%%%%%%%%%%%%%%%%%%
\title{Tunneling from an oscillating state in quantum field theory}
\author{Oliver Janssen}
\affiliation{Laboratory for Theoretical Fundamental Physics, EPFL, 1015 Lausanne, Switzerland}
\affiliation{Institute for Theoretical Physics, KU Leuven, 3001 Leuven, Belgium}
\affiliation{Leuven Gravity Institute, KU Leuven, 3001 Leuven, Belgium}
\author{Matthew Kleban}
\affiliation{Center for Cosmology and Particle Physics, New York University, New York, NY 10003, USA}
\affiliation{Center for Computational Astrophysics, New York, NY 10010, USA}
\author{Mehrdad Mirbabayi}
\affiliation{International Centre for Theoretical Physics, Strada Costiera 11, 34151 Trieste, Italy}
\author{Cameron Norton}
\affiliation{Center for Cosmology and Particle Physics, New York University, New York, NY 10003, USA}
\date{\today}

%%%%%%%%%%%%%%%%%%%%%%%%
\begin{abstract}
How does a scalar field oscillating about a metastable minimum decay? For a coherent state with $\langle q(t) \rangle = a \cos mt$ we find two results. When $mR \ll 1$, with $m$ the mass and $R$ the bubble radius, the decay at time $t$ is a Coleman bubble in the frozen potential $U_t(\eta) = U(\langle q \rangle +\eta)-U(\langle q \rangle)-U'(\langle q \rangle)\eta$, for any amplitude. At small amplitude but any $mR$, $B(t)=B_0 - a \mathcal{M} \cos mt+O(a^2)$, with $\mathcal{M}$ set by the tail of the vacuum bounce; for a thin wall $\mathcal{M} \propto e^{mR}$, so exponentially small oscillations change the rate significantly. Both results follow from a Euclidean source construction. The same method computes the tunneling rate of a field slowly rolling in one direction and metastable in another.
\end{abstract}

\maketitle

%%%%%%%%%%%%%%%%%%%%%%%%
\emph{Introduction.}---A scalar in a metastable vacuum decays by nucleating bubbles of a lower-energy phase, at a rate $\Gamma\propto\e^{-B_0}$ set by Coleman's bubble instanton~\cite{Coleman}. Cosmological misalignment or a resonant drive can instead leave the field oscillating about the minimum, as in metastable axion landscapes~\cite{AxionLandscape}; related transitions occur on rolling inflationary backgrounds~\cite{KlebanNorton,Draper}. The same situation could be engineered in the laboratory. Bubble nucleation from a metastable magnetization has been observed in coherently coupled atomic superfluids~\cite{Zenesini}, where a quench of the detuning would set the field oscillating about the false vacuum, and in long Josephson junctions the phase is a sine-Gordon field whose decay is stimulated by microwaves~\cite{SSN}.

Even an oscillation too small to cross the barrier changes the state from which tunneling starts. Tunneling probes the wavefunction's tail under the barrier, and excited-state tails are much larger than the ground-state tail. Hence an exponentially small excited component can strongly affect decay while barely changing the state near the minimum. Specifying the mean motion therefore does not determine the decay rate. Here we will compute $\Gamma(t)\propto\e^{-B(t)}$ for a gradually prepared coherent state with $\langle q(t)\rangle = a\cos mt$, where at $t=0$ the field is at the classical turning point toward the barrier.

Two complementary limits organize the analysis. In the adiabatic limit when the background varies slowly across a bubble of radius $R$, i.e.~when $mR \ll 1$, we need not expand in its displacement. At small oscillation amplitude $a$, we need not restrict $mR$. A Euclidean source construction provides a common framework. We first explain the mechanism in quantum mechanics, where the source prescription can be checked directly, and then extend it to field theory. We compute exponents only, leaving the semiclassical prefactor to future work.

Despite the fundamental nature of this question, very few results on field theory decay from oscillating states are available in the literature.  The analyses we are aware of~\cite{KK,DJL1,DJL2} disagree on the size of the enhancement but find leading oscillatory corrections to $B(t)$ proportional to $a^2\cos 2mt$. This is even-in-$a$ and does not distinguish oscillations into versus away from the barrier separating the metastable vacuum from the true vacuum. We instead find a leading correction $\Delta B \propto a \cos mt$. Our results in the adiabatic limit agree with \cite{KlebanNorton} and \cite{Yamada}. Recent studies of tunneling from excited states include~\cite{SKK,SteingasserKaiser,GarbrechtWagner,LinSteingasser,LinSteingasserLong,Janssen:2026ybl,JKN}. Computations of decay in the presence of time-dependent \emph{classical} backgrounds include~\cite{BrezinItzykson,DunneSchubert,DASM}.

\begin{figure}[t]
\centering\includegraphics[width=\columnwidth]{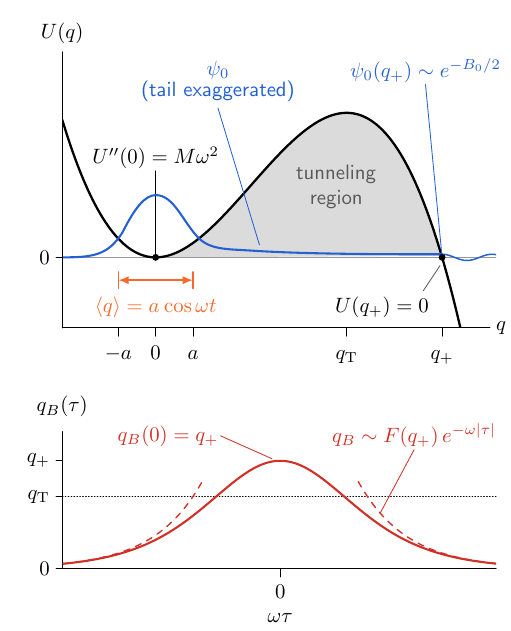}
\caption{\underline{Top}: the metastable well and its ground-state tail (exaggerated). Excited states have larger tails under the barrier. The oscillation amplitude is $a\ll\qT$, and $\psi_0(\qp)\sim\e^{-B_0/2}$. \underline{Bottom}: the zero-energy bounce reaches $\qp$ at $\tau=0$ and approaches the minimum as $F(\qp)\e^{-\omega|\tau|}$ (dashed).}
\label{fig:well}
\end{figure}

%%%%%%%%%%%%%%%%%%%%%%%%
\emph{Quantum-mechanical mechanism.}---Consider a particle of mass $M$ in a well $U$ with a metastable minimum at $q=0$, $U(0)=0$, $U''(0)=M\omega^2$, barrier top at $\qT$ and exit point $\qp$, where $U(\qp)=0$ (Fig.~\ref{fig:well}). Let $\psi_n$ be long-lived resonant states, normalized in the well, with energies $E_n-i\hbar\Gamma_n/2$ and $E_n \approx \hbar\omega (n+1/2)$~\cite{gamow1928, siegert1939, landau1991, JKN}. Here $\Gamma_n$ is the decay rate and $\hbar\Gamma_n$ the energy width. We consider the coherent state
\begin{align}
\psi_a = \sum_n c_n \psi_n \,, \quad c_n = \e^{-\bar{n}/2} \frac{\alpha^n}{\sqrt{n!}} \,, \quad \alpha \equiv \frac{a}{\sqrt{2}d} \,,
\label{eq:coh}
\end{align}
with $\bar n \equiv \alpha^2$, mean motion $\langle q(t)\rangle=a\cos \omega t$, and Poisson weights $P_n=\e^{-\bar n}\bar n^n/n!$. $d \equiv \sqrt{\hbar/(M\omega)}$ is the width of the metastable ground state.

To compute the tunneling rate, we can consider the relative amplitudes of these states at $\qp$. Their enhancement relative to $\psi_0$ is controlled by a coefficient $F$, determined by the asymptotic tail of the zero-energy bounce. Choosing $q_B(0)=\qp$, we define $F(\qp)$ by $q_B(\tau) \sim F(\qp)\e^{-\omega|\tau|}$ as $|\tau| \to \infty$. In a harmonic well $F=\qp$. For a general barrier, using conservation of energy, we find $F/\qp=\exp\int_0^{\qp}(\omega/v-1/q)\,\dd q$ with $v=\sqrt{2U/M}$ (End Matter): $\log(F/\qp)$ is $\omega$ times the excess Euclidean time the bounce spends crossing the barrier compared with a harmonic path. A bounce that lingers under the barrier makes $F/\qp$ exponentially large, the quantum-mechanical counterpart of the thin-wall factor $\e^{mR}$ below~\cite{NoteJosephson}.

For $E_n\ll U(\qT)$ (End Matter and~\cite{Garg2}), $\psi_n(q_+) \simeq \frac{(\sqrt{2}F/d)^n}{\sqrt{n!}} \psi_0(q_+)$. Each resonant-state amplitude acquires the relative phase $e^{-i n\omega t}$, so the coherent sum gives
\begin{equation} \label{eq:x}
\frac{\psi_a(\qp,t)}{\psi_0(\qp,t)} \simeq \exp\!\left( -\frac{\bar n}{2}+\frac{x}{2}e^{-i \omega t}\right) \,, \quad x \equiv 2 \frac{a F(q_+)}{d^2} \,.
\end{equation}
For $x\gg1$, the level dominating the period-averaged decay is $n_*=x/2$, linear in $a$ and potentially far above $\bar n$ (the spectral derivation and the period average are found in the End Matter). At leading exponential order the outgoing rate at time $t$ follows $|\psi_a(\qp, t)|^2$, yielding
\begin{equation}
B(t)=B_0 - 2\frac{a F(\qp)}{d^2} \cos \omega t + O(a^2) \,.
\label{eq:qmlin}
\end{equation}
The rate is largest at the turning point $t=0$, where the field is displaced toward the barrier, and smallest one half period later. The normalization contributes $+\bar n$ to $B$ (one of the omitted $O(a^2)$ terms). The quadratic approximation must hold at the dominant level, $E_* = \hbar\omega n_*\ll U(\qT)$, or $a\ll U(\qT)/[M\omega^2F(\qp)]$.

%%%%%%%%%%%%%%%%%%%%%%%%
\paragraph{State preparation.}
Eq.~\eqref{eq:coh} specifies the state through its resonant-state coefficients $c_n$, and the decay rate depends on them mainly for $n\approx n_*$. In a harmonic well $\hat q$ connects $\psi_m$ only to $\psi_{m\pm1}$, and any preparation that exponentiates an operator linear in $\hat q$ and $\hat p$---displacing the ground state by $a$, multiplying it by $\e^{M\omega a\hat q/\hbar}$, or driving it in real time at frequency $\omega$---yields exactly the coefficients of Eq.~\eqref{eq:coh}. In an anharmonic well $\langle n|\hat q|0\rangle\neq0$ for $n\ge2$. These matrix elements are suppressed by $(d/\qT)^{n-1}$ relative to $\langle1|\hat q|0\rangle$~\cite{BLST,Bachas} and barely affect $\langle q(t)\rangle$ or the state near the minimum, but at $\qp$ they are multiplied by the tail ratios $(\sqrt2F/d)^n/\sqrt{n!}$ and change $B$ at $O(a)$. A ``sudden'' preparation, applied over times $\lesssim1/\omega$, has Fourier power at the transition frequencies $(E_n-E_0)/\hbar$, $n\ge2$, and excites these admixtures; a drive at $\omega$ switched on slowly compared with $1/\omega$ does not. All such gradually prepared oscillations give Eq.~\eqref{eq:qmlin} at $O(a)$.

A convenient gradual preparation that carries over to field theory places a linear source at finite Euclidean depth $\tau_c$:
\begin{equation}
\psi_{\tau_c}\propto\e^{-(H-E_0)\tau_c/\hbar}\,\e^{\lambda\hat q/\hbar}\,\psi_0 \,, \qquad \lambda = M\omega a \, \e^{\omega\tau_c} \,.
\label{eq:prep}
\end{equation}
In a harmonic well this prepares a coherent state with amplitude $a$ for any $\tau_c$, whose dependence drops out after normalizing. In an anharmonic well the source must lie well into the bounce's harmonic tail, $\omega(\tau_c-\tau_B)\gg1$ (where $\e^{\omega\tau_B}\equiv F/\qp$ measures the Euclidean time the bounce spends under the barrier), but remain weak, $a \, e^{\omega\tau_c}\ll\qT$. A natural choice is $\tau_c=\tau_*$, $a \, e^{\omega\tau_*}=\sqrt{2aF}$: the source brings the particle to rest at the classical turning point of level $n_*$, and between the planes the saddle is the Euclidean orbit of energy $E_*$ (End Matter). This is the same trajectory as the complex-time saddle of~\cite{JKN}, whose Euclidean segment runs from $x_L=\sqrt{2aF}$ through the barrier.

%%%%%%%%%%%%%%%%%%%%%%%%
\paragraph{Source method.}
We use complex time $t_{\rm C}=-i\tau$, so the ket contour $\tau:-\infty\to0$ runs from $t_{\rm C}=+i\infty$ to $0$. The ket $\psi_J(Q)=\int_{q(-\infty)=0}^{q(0)=Q}\mathcal{D}q\,\e^{-S_J/\hbar}$ runs from the false vacuum at $\tau\to-\infty$, through the source at $-\tau_c$, to $q(0)=Q$. Reflection and complex conjugation give the bra, with the source at $+\tau_c$. Their product $|\psi_J(Q)|^2$ is a full-line path integral with $q(0)=Q$~\cite{NoteRelatedMethods}. For nucleation at time $t$ we take $\lambda=M\omega a\,\e^{\omega\tau_c-i\omega t}$.
The sourced action is
\begin{equation}
\begin{aligned}
S_J&=\int\!\dd\tau\,
\left( \tfrac12M\dot q^2+U(q)-Jq \right) \,, \\
J(\tau)&=\lambda\delta(\tau+\tau_c) + \lambda^*\delta(\tau-\tau_c) \,.
\end{aligned}
\label{eq:SJ}
\end{equation}
Rotating the source phase implements real-time evolution exactly in a harmonic well and at $O(a)$ within the anharmonic preparation window.

\begin{figure}[t]
\noindent\makebox[\columnwidth][l]{%
  \hspace*{-7pt}%
  \includegraphics[width=\columnwidth]{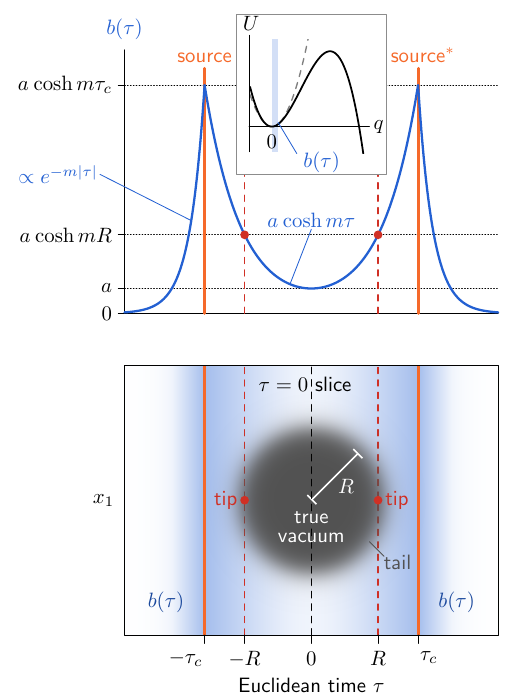}%
}
\caption{The small-amplitude source construction at the favorable turning point $t=0$. \underline{Top}: between the sources, $b=a\cosh m\tau$; beyond them $b$ decays. At the bounce tips, $\tau=\pm R$, $b \simeq a \, \e^{mR}/2$. The inset shows the harmonic range of the background. \underline{Bottom}: the localized bubble sits on this background; its central slice is at $\tau=0$. Shading shows $b$. In quantum mechanics, $m\to\omega$.}
\label{fig:schematic}
\end{figure}

The norm of the state is the full-line path integral with $Q$ integrated over and is dominated by the background saddle $b$. Dividing by the norm gives the probability density an exponent $(S_J[q_Q]-S_J[b])/\hbar$, where $q_Q(0)=Q$. Let $b(t)=\langle q(t)\rangle$ denote the real-time orbit. Between the source planes, the background saddle is its continuation to complex time, $b(t-i\tau)=a\cos[\omega(t-i\tau)]+O(a^2)$. At $t=0$ this is the real, even function $a\cosh\omega\tau$. Beyond the source planes the background turns onto decaying tails and approaches the false vacuum as $|\tau|\to\infty$ (Fig.~\ref{fig:schematic}).

At $t=0$ the bounce is the real, even saddle that reduces to Coleman's as $a \to 0$.  Its relation to real-time decay is checked numerically below. Writing $q=\ba+\eta$, the action difference between bounce and background is exactly
\begin{equation}
\begin{aligned}
 \hbar B &= S_J[\ba+\eta]-S_J[\ba] \\
 &=\int\!\dd\tau\Big( \tfrac12M\dot\eta^2+U(\ba+\eta) -U(\ba)-U'(\ba)\eta \Big) \,.
\end{aligned}
\label{eq:Stilde}
\end{equation}
The cross-term $\int M\dot\ba\dot\eta$ is integrated by parts, the equation of motion of $\ba$, source included, cancels the $J\eta$ term, and the surface term vanishes because $\ba$ decays beyond the sources. The result is an instanton problem for  $\eta$ in the time-dependent (and complexified) potential $U(b+\eta)-U(b)-U'(b)\eta$. 

The linear response is $\dd B/\dd a$ at $a=0$, where the saddle is $\eta=q_B$. Since the saddle is stationary, the derivative can be taken at fixed $\eta$, giving $-\hbar^{-1} \int\dd\tau\,J_B\,\partial_ab$, where $J_B\equiv M\omega^2q_B-U'(q_B)=M(\omega^2-\partial_\tau^2)q_B$ is the source that would produce the bounce in a harmonic well. Between the source planes, $\partial_ab|_{a=0}=\cos\omega t\cosh\omega\tau+i\sin\omega t\sinh\omega\tau$. The odd part integrates to zero against the even $J_B$ and
\begin{equation}
 B(t) = B_0-\frac{a\M}{\hbar}\cos\omega t + O(a^2) \,,~ \M = \int\!\dd\tau\,J_B\,\e^{\omega\tau} \,,
\label{eq:qmM}
\end{equation}
where $\cosh\omega\tau$ has been replaced by $\e^{\omega\tau}$ using the symmetry of $q_B$. Integrating by parts leaves only the boundary at $\tau\to+\infty$, where $q_B \sim F(\qp)\e^{-\omega\tau}$: $\M=2M\omega F(\qp)$, and Eq.~\eqref{eq:qmM} reproduces Eq.~\eqref{eq:qmlin}. The source method therefore agrees with the coherent-state calculation at $O(a)$.

%%%%%%%%%%%%%%%%%%%%%%%%
\emph{Field theory.}---The source construction carries over with $\int\dd\tau\to\int\dd^4x$, $M\to1$, and $\omega\to m$. We  set $\hbar=1$ hereafter. For a real scalar, $S=\int\dd^4x[\tfrac12(\partial q)^2+U(q)]$, with $U''(0)=m^2$, Euclidean Laplacian $\Box=\partial_\tau^2+\nabla^2$, and $O(4)$-symmetric Coleman vacuum bubble $q_B(\rho)$, $\rho=\sqrt{\tau^2+|\mathbf x|^2}$. The condensate displaces the zero-momentum mode, and Eq.~\eqref{eq:SJ} describes two spatially homogeneous source planes with $\lambda = am \, \e^{m\tau_c-imt}$. The homogeneous solution $b$ decays beyond the planes and is the continued orbit $b(t-i\tau)=a\cos[m(t-i\tau)]+O(a^2)$ between them, just as in quantum mechanics.

At $t=0$ the bubble is a second, real, even saddle, $q=b+\eta(\tau,\mathbf x)$, with $\eta$ localized in space and Euclidean time and reducing to Coleman's bubble as $a\to0$. It satisfies
\begin{equation}
(\partial_\tau^2+\nabla^2)\eta=U'(b+\eta)-U'(b) \,,
\end{equation}
and has exponent $S_J[b+\eta]-S_J[b]$.

%%%%%%%%%%%%%%%%%%%%%%%%
\emph{Frozen background: small $mR$.}---When the background changes little across the bubble, set $b(t-i\tau) = b(t)$ inside the integrand of Eq.~\eqref{eq:Stilde}. The exponent is Coleman's action for fluctuations about a fixed  background potential~\cite{Yamada}:
\begin{equation}
\begin{aligned}
 B(t) &\simeq B_{\rm Col}[U_t] \,, \\
 U_t(\eta)&=U(b(t)+\eta)-U(b(t))-U'(b(t))\eta \,.
\end{aligned}
\label{eq:frozen}
\end{equation}
The term $U'(b(t))\eta$ arises because the background obeys its equation of motion, so the fluctuation action has no linear term and $U_t'(0)=0$. Note that even in infinite volume $B$ is finite because $\eta$ is localized, even though each $S_J$ diverges.

Three conditions make Eq.~\eqref{eq:frozen} meaningful. For a Coleman bounce in $U_t$ to exist, $U_t$ must have a metastable minimum at $\eta=0$, $U_t''(0)=U''(b)>0$, and a region of lower potential beyond its barrier.  These can fail when the background approaches the barrier top, or when it has moved far enough away from it that the lower well of $U_t$ closes. For the frozen bounce to be the saddle of Eq.~\eqref{eq:Stilde}, the background must change little across it. Between the sources $b$ obeys the equation of motion, so $b(t-i\tau)=b(t)-i\tau\dot b+\tfrac12\tau^2U'(b)+\cdots$ for any amplitude. This expansion can be used to systematically compute corrections to the frozen limit and estimate the errors.  The term odd in $\tau$ integrates to zero against the even bounce, so it shifts $B$ only at second order.  The leading corrections are therefore $O((mR)^2)$, and requiring them to shift the exponent by much less than one gives $|\partial_bB|\,|U'(b)|R^2\ll1$ and $|\partial_b^2B|\,(\dot bR)^2\ll1$. For a small harmonic oscillation both reduce to $a|\partial_bB|(mR)^2\ll1$. In infinite volume the homogeneous mode has no quantum spread, so Eq.~\eqref{eq:frozen} is evaluated at the sharp value $b=\langle q(t)\rangle$. In quantum mechanics there is a finite spread and the rate averaged over the distribution exceeds the rate at the mean (see the slow-coordinate test in End Matter). For multiple fields $q^i$ the same analysis gives $U_t(\eta)=U(b+\eta)-U(b)-\partial_iU(b)\,\eta^i$, so a time-dependent spectator field can modulate the tunneling potential of another.

%%%%%%%%%%%%%%%%%%%%%%%%
\emph{Linear response: arbitrary $mR$.}---Instead of freezing $\ba$, expand Eq.~\eqref{eq:Stilde} to first order in $a$ on the vacuum bubble. Stationarity of the saddle eliminates its first-order deformation, so $\Delta B=-\int\dd^4x\,J_B\,b$ with $J_B=(m^2-\Box)q_B$. At this order $b$ obeys $(m^2-\partial_\tau^2)b=J$, and integrating by parts,
\begin{equation}
 \Delta B = -\!\int\!\dd^4x\,q_B\,J = -2am\,\e^{m\tau_c} \, Q_B(\tau_c) \cos mt \,,
\label{eq:response}
\end{equation}
where $Q_B(\tau)=\int\dd^3 \mathbf{x} \,q_B$. This holds for any source depth. Where $J_B$ is negligible, $Q_B$ obeys the free equation, $Q_B(\tau_c) \simeq (\M/2m)\e^{-m|\tau_c|}$, with
\begin{equation}
\M=\int\!\dd^4x\,J_B\,\e^{m\tau} = \tilde J_B(k)\big|_{k^2=-m^2} \,.
\label{eq:lin}
\end{equation}
For $\tau_c-R\gg1/m$ this gives $B(t)=B_0-a\M\cos mt+O(a^2)$, with corrections $O(\e^{-m(\tau_c-R)})$ (for an even potential, $O(\e^{-2 m(\tau_c-R)})$) from the  nonlinear residual of $q_B$ at $\tau_c$. Inside the preparation window these are negligible and $a \, e^{m\tau_c}\ll\qT$ holds.

The coefficient depends only on the vacuum Coleman bubble. It makes contact with the frozen result: by reflection symmetry $\M=\int J_B\left(1+\tfrac12m^2\tau^2+\cdots\right)$, whose first term is the frozen-background coefficient, the rest measuring the effect of larger $mR$. As in quantum mechanics, $\M$ is a tail coefficient, just as $q_B \simeq F\e^{-\omega|\tau|}$ with $\M=2M\omega F$. Equivalently, $\tilde q_B=\tilde J_B/(k^2+m^2)$, so $\M$ is the residue of $\tilde q_B$ at the one-particle pole: the amplitude for the bubble to absorb one on-shell quantum. Since the condensate is a coherent state of zero-momentum quanta, the linear response is single-quantum absorption. The same residue governs decay induced by a single particle~\cite{GorskyVoloshin}.

Write $x=a\M$. Keeping the linear term in the exponent gives $\Gamma(t) \simeq \Gamma_0\e^{x\cos mt}$ and, averaged over a period, $\bar\Gamma/\Gamma_0 \simeq I_0(x)\sim\e^x/\sqrt{2\pi x}$ for $x\gg1$. This resums the linear exponent but not all the $O(a^2)$ effects. It is useful when $x$ is appreciable but omitted terms in $B$ remain small.

%%%%%%%%%%%%%%%%%%%%%%%%
\emph{Large $mR$.}---In the thin-wall regime, outside the wall the bubble has the free tail $q_B \simeq c\,\e^{-m(\rho-R)}(R/\rho)^{3/2}$, where $R$ is the radius at which $q_B=\qT$ and $c=\qT\exp\int_0^{\qT}\dd q\,(m/\sqrt{2U}-1/q)$, the integral running along the one-dimensional wall profile. This is the field-theory analogue of $F$, Eq.~\eqref{eq:F}, with $\qp\to\qT$; $c$ is of order $\qT$ ($c=0.86\,\qT$ in our numerical example).

The spatial integral of $q_B$ at $|\tau|\gg R$ is $c\,(2\pi R/m)^{3/2}\e^{mR}\e^{-m|\tau|}$, so Eqs.~\eqref{eq:response} and \eqref{eq:lin} give
\begin{equation}
 B(t) = B_0 - 2 \frac{ac}{m^2} (2\pi m R)^{3/2} \e^{mR}\cos mt + O(a^2) \,.
\label{eq:Bthin}
\end{equation}
The exponential comes from the bubble's Euclidean-time tips at $\tau \approx \pm R$, where the continued background $b(-i\tau)$ reaches $a\cosh mR$ (Fig.~\ref{fig:schematic}). Consequently, an oscillation with exponentially small amplitude can change $B$ by order one when $mR \gg 1$.

Apart from the frozen regime $mR \ll 1$, the decay rate at large $a$ remains unresolved. In our large-$mR$ numerical example, the real saddle connected to Coleman's bounce merges with a second branch near $x \simeq 40$. The fold resembles that in thin-wall thermal decay~\cite{Garriga}. We leave the question of which saddles contribute to future work.

%%%%%%%%%%%%%%%%%%%%%%%%
\emph{Energy of the first decays.}---Quantum-mechanical decays sample $P_n\Gamma_n$. Before significant depletion when the $P_n$ are Poisson, the typical excitation energy $E_* = \hbar\omega n_* $ exceeds the initial mean $\langle H\rangle-E_0 = \hbar\omega \bar n$ by $n_*/\bar n=2F/a$. The field-theory counterpart is the turning-point bubble's energy relative to the homogeneous background (End Matter),
\begin{equation}
 E_{\rm bub} = \tfrac12ma\M+O(a^2) \,.
 \label{eq:bubble-energy}
\end{equation}
At large $mR$, the exponential growth of $\M$ lets this greatly exceed the mean condensate energy in the bubble volume.  Nevertheless energy is conserved: the energy density of the undecayed part of the condensate decreases as its resonant components damp like $e^{-\Gamma t}$, which (on average) cancels the higher energy of the bubbles. The oscillating state itself has a finite lifetime, polynomial in the parameters, from parametric resonance of the non-zero modes and subsequent thermalization~\cite{AminReview}; our rate applies during that window.

%%%%%%%%%%%%%%%%%%%%%%%%
\emph{Relation to earlier work.}---Keski-Vakkuri and Kraus~\cite{KK} obtain a correction $\propto a^2\e^{2mR}\cos2mt$ from a membrane action in an oscillating background. An ideal thin membrane has no exponential tail, and it is the tail that produces the leading term linear in $a$. Darm\'e, Jaeckel and Lewicki first froze the exterior oscillation~\cite{DJL1}, and later treated its phase as a real collective coordinate in a WKB analysis on the $(R,T)$ hypersurface~\cite{DJL2}.  In both, the displacement enters quadratically, through the exterior energy density and pressure, with the wall tension and interior unchanged, so the result is even in $a$. Because the phase is never continued into Euclidean time, the exponential growth of the background across the bounce (which controls the answer at large $mR$) is absent from their treatment. A quadratic response is the correct leading term when the oscillating field is a $\mathbb Z_2$-odd spectator, which cannot be absorbed one quantum at a time. Ref.~\cite{CMR} finds $\delta B\propto a^2\cos2\omega t$ for this case, a coherent worldsheet-axion excitation on a breaking QCD string. Yamada~\cite{Yamada} also considered tunneling in slowly rolling multifield backgrounds; our results are in agreement.

%%%%%%%%%%%%%%%%%%%%%%%%
\emph{Numerical checks.}---In quantum mechanics we test against exact time evolution. In QFT we can only test the internal consistency of the sourced saddle; a simulation of a 1+1-dimensional field theory could provide an exact test. In the cubic well in QM, resonance tails and independently propagated currents approach Eq.~\eqref{eq:qmlin} to $0.2\%$ as $\hbar$ decreases. A second, slow coordinate tests the frozen limit: four smooth preparations agree with the static rate. In QFT, lattice solutions of the sourced saddle in $D=4$ reproduce $\M$ to $10^{-4}$ and follow the saddle to $x \simeq 30$, where nonlinear corrections reach $15\%$. The End Matter gives representative parameters and convergence data.

\vspace{0.5cm}
%%%%%%%%%%%%%%%%%%%%%%%%
\begin{acknowledgments}
We thank David Cyncynates, Luc Darm\'e, Joerg Jaeckel, David Kaiser, Toby Opferkuch, Rudin Petrossian-Byrne, Massimo Porrati, and Giovanni Villadoro  for useful discussions, and the Bernoulli Center for Fundamental Studies (EPFL, Lausanne) for hospitality during the workshop ``\href{https://indico.epfl.ch/event/5/}{Observers, wormholes, and complex saddles in cosmology}''. OJ is supported in part by the Belgian Federal Science Policy Office (BELSPO) in the framework of the PRODEX Program of the European Space Agency (ESA) under contract number PEA4000144253, and by KU Leuven under grant C16/25/010 ``Holography on the Horizon.''  MK and CN are supported by NSF grant PHY-2412899.
\end{acknowledgments}

%%%%%%%%%%%%%%%%%%%%%%%%

\clearpage
%%%%%%%%%%%%%%%%%%%%%%%%
\section*{End Matter}
\appendix

%%%%%%%%%%%%%%%%%%%%%%%%
\section{Resonant-state tails and the level sum} \label{app:resonances}
We can find the decay of $\psi_n$ in the barrier from WKB (see~\cite{JKN} for a more elaborate discussion). Away from the turning points, where $E_n\ll U(q)$, with $v(q)=\sqrt{2U(q)/M}$,
\begin{equation}
\frac{\dd}{\dd q}\log\frac{\psi_n}{\psi_0}
\simeq \frac{E_n-E_0}{\hbar v(q)} \simeq \frac{n\omega}{v(q)} \,.
\end{equation}
Integrating measures the Euclidean travel time. Matching to the harmonic wavefunctions fixes the normalization and gives
$\psi_n(\qp)/\psi_0(\qp) \simeq (\sqrt{2} F/d)^n/\sqrt{n!}$, with
\begin{equation}
\frac{F}{\qp} = \exp\!\int_0^{\qp}\!\left(\frac{\omega}{v}-\frac1q\right)\dd q \,.
\label{eq:F}
\end{equation}
The subtraction removes the logarithm. This is the same $F$ as what multiplies the exponential tail of the bounce. Before depletion of the relevant levels, the period-averaged rate is
\begin{equation}
\bar\Gamma=\sum_nP_n\Gamma_n \,, \qquad \Gamma_n\propto\e^{-W(E_n)/\hbar} \,.
\label{eq:sum}
\end{equation}
Here $-\dd W/\dd E=2\tau(E)$, where $\tau(E)$ is the Euclidean traversal time between $U(q_{1,2})=E$. Extremizing over $n$ assuming $n \gg 1$ gives
\begin{equation}
n_*=\bar n\,\e^{2\omega\tau(E_*)} \,, \qquad E_* = \hbar\omega n_* \,.
\label{eq:nstar}
\end{equation}
Thus a long traversal selects rare levels far above the mean. At low energy,
\begin{equation}
\e^{\omega\tau(E)} \simeq F\sqrt{\frac{2M\omega^2}{E}} \,, \qquad n_* = \frac{aF}{d^2} = \frac{x}{2} \,.
\label{eq:nstar2}
\end{equation}
This requires $n_*\gg1$ and $E_*\ll U(\qT)$. The mean excitation energy $E_a = M \omega^2 a^2/2$ is quadratic in $a$, whereas $E_*=M\omega^2aF \gg E_a$ is linear. We stress again that this describes early-time decays. At later times, the energy of transmitted particles will lie closer to $E_a$. The sum \eqref{eq:sum} can also be done exactly. The tail ratio gives $\Gamma_n/\Gamma_0 \simeq (2F^2/d^2)^n/n!$, so $\bar\Gamma/\Gamma_0 \simeq e^{-\bar n}I_0(x)$, with saddle at $n_*=x/2$. This is the period average of $|\psi_a(\qp,t)/\psi_0(\qp)|^2=\exp(-\bar n+x\cos\omega t)$, the square of the amplitude sum $e^{-\bar{n}/2} \sum_n(x \, \e^{-i\omega t}/2)^n/n!$. Both forms resum the term linear in $a$ in the exponent.  The $O(a^2)$ terms of $B$ other than $+\bar n$ are not included (see~\cite{JKN}).

%%%%%%%%%%%%%%%%%%%%%%%%
\section{Preparation accuracy and source matching} \label{app:preparation}
The coefficients of the state in Eq.~\eqref{eq:prep} obey
\begin{equation}
\frac{c_n}{c_0} = \e^{-(E_n-E_0)\tau_c/\hbar} \frac{\langle n|\e^{\lambda\hat q/\hbar}|0\rangle}{\langle0|\e^{\lambda\hat q/\hbar}|0\rangle} \,.
\label{eq:coefficients}
\end{equation}
To compare this state with \eqref{eq:coh}, calibrate $\alpha=c_1/c_0$. For our purposes accuracy requires $\sqrt{n!}\,c_n/(c_0\alpha^n)$ close to one over the levels near $n_*$. For a harmonic well the ratios are exactly Poisson. In an anharmonic well, Eq.~\eqref{eq:qmM} gives the finite-depth coefficient $\M(\tau_c)=2M\omega \, \e^{\omega\tau_c}q_B(\tau_c)$. The cubic $U=q^2/2-q^3/3$, in units $M=\omega=1$, makes this explicit. The coherent-state result, Eq.~\eqref{eq:qmlin}, has $\M=2M\omega F=12$, while
\begin{equation}
q_B = \tfrac32\operatorname{sech}^2(\tau/2) \,, \qquad \M(\tau_c)=\frac{12}{(1+\e^{-\tau_c})^2} \,.
\label{eq:cubic-depth}
\end{equation}
A ``sudden" source at $\tau_c=0$ gives $3$, while $\M$ converges to the coherent-state value exponentially at large $\tau_c$. Increasing depth also changes the desired coefficients through the anharmonic gaps. After absorbing the term linear in $n$ into $\alpha$, let
\begin{equation}
\delta E_n = E_n - E_0 - n(E_1-E_0) \simeq \tfrac12\hbar n(n-1)\frac{\dd\omega}{\dd n} \,,
\end{equation}
where $\dd\omega/\dd n\sim\omega/n_{\max}$ is the frequency shift per quantum and $n_{\max}\simeq U(\qT)/\hbar\omega$ the number of levels in the well ($\dd\omega/\dd n=-5\hbar/6$ for the cubic above). The extra factor $\e^{-\delta E_n\tau_c/\hbar}$ changes the favorable-phase exponent by $n_*^2\tau_c\,\dd\omega/\dd n$ at leading order.

There are two natural notions of accuracy to $O(a^2)$. For the $O(a)$ coefficient to be correct we need the $O(a^2)$ term to be much less than $x$:
\begin{equation}
\left|\frac{\delta\M}{\M}\right| \ll 1 \,, \qquad n_*\tau_c\left|\frac{\dd\omega}{\dd n}\right|\sim\frac{n_*}{n_{\max}}\,\omega\tau_c \ll 1 \,,
\label{eq:prep-rel}
\end{equation}
i.e.\ $\omega(\tau_c-\tau_B)\gg1$ and a mild bound on $n_*$. For $B$ itself to be accurate to $O(1)$ is more stringent.  The $O(a^2)$ terms must be small compared to one,
\begin{equation}
x\left|\frac{\delta\M}{\M}\right| \ll 1 \,, \qquad
n_*^2\tau_c\left|\frac{\dd\omega}{\dd n}\right|\sim\bar n\,\omega\tau_c\Big(\frac{F}{\qT}\Big)^{2} \ll 1 \,,
\label{eq:prep-errors}
\end{equation}
using $n_*^2\sim(F/\qT)^2n_{\max}\bar n$. The second requires $\bar n\lesssim1$, an amplitude below the zero-point length; this is compatible with $n_*\gg1$ since $n_*/\bar n=2F/a$. Both hold in addition to $a \, e^{\omega\tau_c}\ll\qT$. In real time the corresponding factors $\e^{-i\delta E_nt/\hbar}$ are phases that drop out of $|\psi|^2$ at $n_*$; only the levels within the saddle width $\sqrt{n_*}$ dephase, after $t\sim n_{\max}/(n_*\omega)$, many periods. In field theory the same effect is negligible: the zero mode’s level spacing departs from harmonic by $\hbar\delta\omega_n=E_n-E_0-n(E_1-E_0)$, and at the level $n_*=a\M/2$ selected by the decay, which is independent of the volume $\mathcal V$, $\delta\omega_{n_*}\sim m\,a_*^2/\qT^2$ with $a_*^2=2n_*\hbar/\mathcal Vm$ the classical amplitude of that level.  The resulting shift $n_*\tau_c\delta\omega_{n_*}$ of the exponent vanishes as $1/\mathcal V$.

Within the window a natural depth is $\tau_*=\tau(E_*)$ of Eq.~\eqref{eq:nstar2}, i.e.\ $a \, e^{\omega\tau_*}=\sqrt{2aF}$. The source first builds a coherent state of mean occupation $\bar n_s=\bar n\,\e^{2\omega\tau_c}$, which the Euclidean evolution damps back to $\bar n$ while preserving the ratios of the $c_n$; the choice of $\tau_c$ is therefore the choice of which level sits at the center of the state when the kick is applied.  At $\tau_c=\tau_*$ one has $\bar n_s=n_*$: the source displaces the oscillator to the classical turning point of the dominant level, $\sqrt{2aF}=\sqrt{2E_*/M\omega^2}$, and the background $a\cosh\omega\tau_*$ equals the bounce tail $F\e^{-\omega\tau_*}$, both $\simeq \sqrt{aF/2}$. The window exists when this lies below the barrier, $aF\ll\qT^2$.

The matching in the path integral is elementary: $M\ddot q=U'(q)-J$ gives jumps $-\lambda/M$ and $-\lambda^*/M$ at $\mp\tau_c$. At $t=0$ the harmonic background is $a\cosh\omega\tau$ between the planes and $a\cosh(\omega\tau_c)\e^{-\omega(|\tau|-\tau_c)}$ outside. Rotating the source evolves a harmonic state exactly. For the filtered anharmonic state it gives the linear response up to finite-depth admixtures.

The $O(a^2)$ terms of $B$ depend on the preparation. In the Euclidean preparation they contain products of the growing background with the bounce tail, $b^2q_B^2\sim a^2$ at each $\tau$, integrated up to the source. In quantum mechanics this gives a term growing linearly in $\tau_c$: the anharmonic-gap effect above. In $D$ spacetime dimensions the transverse spread of the tail over distances $\sqrt{|\tau|/m}$ gives $\int\dd^{D-1}x\,b^2q_B^2\propto|\tau|^{-(D-1)/2}$, whose depth integral converges for $D\ge4$. In four dimensions the $O(a^2)$ terms therefore do not grow with depth, though they remain preparation-dependent, as they must.

%%%%%%%%%%%%%%%%%%%%%%%%
\section{Energy of the bubble} \label{app:energy}
At the favorable turning point, $\dot q=\dot b=0$. The bubble's classical energy relative to the homogeneous background is
\begin{equation}
E_{\rm bub} = \int\dd^3 \mathbf{x}\,\left[ \tfrac12(\nabla q)^2+U(q)-U(b) \right] \,.
\end{equation}
The relative Euclidean Hamiltonian is constant between the sources, vanishes beyond them, and equals $-E_{\rm bub}$ at the midpoint. Its jump at $+\tau_c$ therefore determines
\begin{equation}
E_{\rm bub} = -\lambda\int\dd^3 \mathbf{x} \, \partial_\tau\eta(\tau_c,\mathbf x)
\simeq \tfrac12ma\M \,,
\end{equation}
where the last step uses $\eta \simeq q_B$, $Q_B \simeq (\M/2m)\e^{-m\tau}$, and $\lambda = am \, \e^{m\tau_c}$. This energy is linear in $a$, unlike the mean condensate energy density $\rho_{\rm osc}=m^2a^2/2$. Like a sphaleron in thermal field theory, the bubble has higher energy density than the condensate it nucleated from. This is possible because the bubble is the most probable decay from a superposition of energy eigenstates.

%%%%%%%%%%%%%%%%%%%%%%%%
\section{Numerical checks}
\label{app:numerics}

%%%%%%%%%%%%%%%%%%%%%%%%
\paragraph{Linear coefficient in QM.}
We prepare the state of Eq.~\eqref{eq:prep} in the cubic well $U=q^2/2-q^3/3$ ($M=\omega=1$) and measure the outgoing current $j$ at a detector beyond the barrier in two independent ways: direct real-time propagation with an absorber, and evolution of the resonant-state expansion of the same saved state. The two agree in current to about $10^{-5}$. At $\hbar=0.03$, $\tau_c=3.6$, $a=0.002$, detector $q=2.5$, halving the spatial step changes the current by $0.13\%$; moving the absorber onset from $2.8$ to $3.0$ changes it by $0.087\%$. The quantity compared with Eq.~\eqref{eq:qmlin} is the fundamental Fourier component of $\hbar\log(j/P_{\rm well})$, with $P_{\rm well}$ the surviving probability in the well, differentiated with respect to the source amplitude and normalized to the predicted $\M=12$ (Table~\ref{tab:current}). Its frequency is the measured level spacing. Both depths extrapolate linearly in $\hbar$ to $1.00$ (to $0.2\%$); the residual $\tau_c$ dependence is a finite-$\hbar$ effect. At these parameters $x=2M\omega aF/\hbar \simeq 0.8$ and $\bar n \simeq 7 \times 10^{-5}$. 

\begin{table}[h]
\centering
\begin{tabular}{@{}lccc@{}}
\toprule
$\hbar$ & $0.050$ & $0.035$ & $0.030$\\
\midrule
Fundamental$/12$, $\tau_c=3.6$ & $1.0455$ & $1.0326$ & $1.0280$\\
Fundamental$/12$, $\tau_c=4.0$ & $1.0662$ & $1.0459$ & $1.0392$\\
\bottomrule
\end{tabular}
\caption{Cubic current response, using the source amplitude of Eq.~\eqref{eq:prep}. The smaller-$\hbar$ propagations did not pass the current-precision checks and are omitted.}
\label{tab:current}
\end{table}

%%%%%%%%%%%%%%%%%%%%%%%%
\paragraph{Freezing in QM.}
A cubic tunneling coordinate $q$ is coupled by $\kappa qy$ to a slow oscillator $y$ of frequency $\Omega$, so that $y$ plays the role of a slowly varying background. At $\Omega/\omega=0.02$, four smooth preparations of $y$ agree to $0.5\%$ in $\log\Gamma$ with the static rate averaged over the distribution of $y$, $\int\dd y\,\rho_y(y,t)\Gamma(\kappa y)$, where $\Gamma(\kappa y)$ is the rate at fixed $y$.

%%%%%%%%%%%%%%%%%%%%%%%%
\paragraph{Sourced saddle in $D=4$.}
This tests Eq.~\eqref{eq:response} and the continuation of the saddle to finite $a$, not real-time decay. We solve the $O(3)$ saddle equation on the $(\tau,r)$ half-plane for $U=\tfrac12q^2+\tfrac{0.25}{4}q^4-\tfrac{3.84}{6}q^6+\tfrac{1.90959}{8}q^8$, with $m=1$, $R=15.013$ (barrier crossing), $B_0=10922.9$, and source planes at $\tau_c=17.092$. The potential is even; we quote the bubble toward the displaced side, which dominates. On $181^2$, $361^2$, and $721^2$ grids of extent $25.01$, centered differences of $\Delta B$ between small sources of opposite sign give the slopes $-\partial\Delta B/\partial b(0)=(4.955,5.042,5.063)\times10^{9}$, extrapolating to $5.0704\times10^{9}$. The prediction of Eq.~\eqref{eq:response} from the vacuum bounce at this depth is $\M(\tau_c)=5.0705\times10^{9}$, and $\M=\tilde J_B(k)|_{k^2=-m^2}=5.0703\times10^9$ at infinite depth; Eq.~\eqref{eq:Bthin} with $c=0.863$ gives $5.24\times10^{9}$. Continuing to a finite source, $\lambda=0.1609$, corresponding to $a \simeq 6.08\times10^{-9}$ and $x=a\M \simeq 31$, the three grids give $\Delta B=B-B_0=-34.86$, $-35.24$, $-35.33$: the magnitude of the change in the action exceeds the linear prediction by about $15\%$. Enlarging the box at fixed spacing changes $\Delta B$ by $6\times10^{-7}$.


\begin{thebibliography}{99}

\bibitem{Coleman} S.~Coleman, \href{https://doi.org/10.1103/PhysRevD.15.2929}{Phys.\ Rev.\ D \textbf{15}, 2929 (1977)} [Erratum: \href{https://doi.org/10.1103/PhysRevD.16.1248}{Phys.\ Rev.\ D \textbf{16}, 1248 (1977)}]; C.~G.~Callan, Jr., and S.~Coleman, \href{https://doi.org/10.1103/PhysRevD.16.1762}{Phys.\ Rev.\ D \textbf{16}, 1762 (1977)}.

\bibitem{AxionLandscape} T.~C.~Bachlechner, K.~Eckerle, O.~Janssen, and M.~Kleban, \href{https://doi.org/10.1088/1475-7516/2019/09/062}{JCAP \textbf{09}, 062 (2019)} [\href{https://arxiv.org/abs/1810.02822}{arXiv:1810.02822}].

\bibitem{KlebanNorton} M.~Kleban and C.~E.~Norton, \href{https://doi.org/10.1103/PhysRevD.111.023538}{Phys.\ Rev.\ D \textbf{111}, 023538 (2025)} [\href{https://arxiv.org/abs/2310.09898}{arXiv:2310.09898}].

\bibitem{Draper} P.~Draper, M.~Karydas, and H.~Zhang, \href{https://doi.org/10.1103/PhysRevD.108.096038}{Phys.\ Rev.\ D \textbf{108}, 096038 (2023)} [\href{https://arxiv.org/abs/2306.01609}{arXiv:2306.01609}].

\bibitem{Zenesini} A.~Zenesini, A.~Berti, R.~Cominotti, C.~Rogora, I.~G.~Moss, T.~P.~Billam, I.~Carusotto, G.~Lamporesi, A.~Recati, and G.~Ferrari, \href{https://doi.org/10.1038/s41567-023-02345-4}{Nat.\ Phys.\ \textbf{20}, 558 (2024)} [\href{https://arxiv.org/abs/2305.05225}{arXiv:2305.05225}].

\bibitem{SSN} A.~O.~Sboychakov, S.~Savel'ev, and F.~Nori, \href{https://doi.org/10.1103/PhysRevB.78.134518}{Phys.\ Rev.\ B \textbf{78}, 134518 (2008)} [\href{https://arxiv.org/abs/0804.2173}{arXiv:0804.2173}].

\bibitem{IvlevMelnikov} B.~I.~Ivlev and V.~I.~Mel'nikov, \href{https://doi.org/10.1103/PhysRevLett.55.1614}{Phys.\ Rev.\ Lett.\ \textbf{55}, 1614 (1985)}.

\bibitem{Garg2} A.~Edison and A.~Garg, \href{https://doi.org/10.1119/5.0236109}{Am.\ J.\ Phys.\ \textbf{93}, 737 (2025)}.

\bibitem{KK} E.~Keski-Vakkuri and P.~Kraus, \href{https://doi.org/10.1103/PhysRevD.54.7407}{Phys.\ Rev.\ D \textbf{54}, 7407 (1996)} [\href{https://arxiv.org/abs/hep-th/9604151}{arXiv:hep-th/9604151}].

\bibitem{DJL1} L.~Darm\'e, J.~Jaeckel, and M.~Lewicki, \href{https://doi.org/10.1103/PhysRevD.96.056001}{Phys.\ Rev.\ D \textbf{96}, 056001 (2017)} [\href{https://arxiv.org/abs/1704.06445}{arXiv:1704.06445}].

\bibitem{DJL2} L.~Darm\'e, J.~Jaeckel, and M.~Lewicki, \href{https://doi.org/10.1103/PhysRevD.100.096012}{Phys.\ Rev.\ D \textbf{100}, 096012 (2019)} [\href{https://arxiv.org/abs/1907.04865}{arXiv:1907.04865}].


\bibitem{Yamada} M.~Yamada, ``Tunneling from a slowly rolling multifield background using tadpole subtraction,'' \href{https://arxiv.org/abs/2608.20451}{arXiv:2608.20451}.

\bibitem{SKK} T.~Steingasser, M.~K\"onig, and D.~I.~Kaiser, \href{https://doi.org/10.1103/PhysRevD.110.L111902}{Phys.\ Rev.\ D \textbf{110}, L111902 (2024)} [\href{https://arxiv.org/abs/2310.19865}{arXiv:2310.19865}].


\bibitem{SteingasserKaiser} T.~Steingasser and D.~I.~Kaiser, \href{https://doi.org/10.1103/PhysRevD.111.096009}{Phys.\ Rev.\ D \textbf{111}, 096009 (2025)} [\href{https://arxiv.org/abs/2402.00099}{arXiv:2402.00099}].

\bibitem{GarbrechtWagner} B.~Garbrecht and N.~Wagner, \href{https://doi.org/10.1007/JHEP05(2025)076}{JHEP \textbf{05}, 076 (2025)} [\href{https://arxiv.org/abs/2412.20431}{arXiv:2412.20431}].

\bibitem{LinSteingasser} J.~Lin, B.~Scheihing-Hitschfeld, and T.~Steingasser, \href{https://doi.org/10.1103/vlvy-3xgq}{Phys.\ Rev.\ D \textbf{114}, L011903 (2026)} [\href{https://arxiv.org/abs/2511.08669}{arXiv:2511.08669}].

\bibitem{LinSteingasserLong} J.~Lin, B.~Scheihing-Hitschfeld, and T.~Steingasser, ``Quantum tunneling from excited states in the steadyon picture,'' \href{https://arxiv.org/abs/2511.08670}{arXiv:2511.08670}.

\bibitem{Janssen:2026ybl} O.~Janssen, J.~Karlsson, F.~Riccardi, and M.~Varrone, ``An ode to instantons,'' \href{https://arxiv.org/abs/2603.06575}{arXiv:2603.06575}.

\bibitem{JKN} O.~Janssen, M.~Kleban, and C.~Norton, ``Tunneling from an oscillating state in quantum mechanics,'' \href{https://arxiv.org/abs/2605.04139}{arXiv:2605.04139}.



\bibitem{gamow1928} G.~Gamow, \href{https://doi.org/10.1007/BF01343196}{Z.\ Phys.\ \textbf{51}, 204 (1928)}.

\bibitem{siegert1939} A.~J.~F.~Siegert, \href{https://doi.org/10.1103/PhysRev.56.750}{Phys.\ Rev.\ \textbf{56}, 750 (1939)}.

\bibitem{landau1991} L.~D.~Landau and E.~M.~Lifshitz, \textit{Quantum Mechanics: Non-Relativistic Theory}, 3rd ed., Course of Theoretical Physics, Vol.~3 (Butterworth-Heinemann, Oxford, 1991).

\bibitem{AminReview} M.~A.~Amin, M.~P.~Hertzberg, D.~I.~Kaiser, and J.~Karouby, \href{https://doi.org/10.1142/S0218271815300037}{Int.\ J.\ Mod.\ Phys.\ D \textbf{24}, 1530003 (2015)} [\href{https://arxiv.org/abs/1410.3808}{arXiv:1410.3808}].

\bibitem{GorskyVoloshin} A.~Gorsky and M.~B.~Voloshin, \href{https://doi.org/10.1103/PhysRevD.73.025015}{Phys.\ Rev.\ D \textbf{73}, 025015 (2006)} [\href{https://arxiv.org/abs/hep-th/0511095}{arXiv:hep-th/0511095}].

\bibitem{Garriga} J.~Garriga, \href{https://doi.org/10.1103/PhysRevD.49.5497}{Phys.\ Rev.\ D \textbf{49}, 5497 (1994)} [\href{https://arxiv.org/abs/hep-th/9401020}{arXiv:hep-th/9401020}].

\bibitem{DASM} R.~Sch\"utzhold, H.~Gies, and G.~Dunne, \href{https://doi.org/10.1103/PhysRevLett.101.130404}{Phys.\ Rev.\ Lett.\ \textbf{101}, 130404 (2008)} [\href{https://arxiv.org/abs/0807.0754}{arXiv:0807.0754}].

\bibitem{BrezinItzykson} E.~Br\'ezin and C.~Itzykson, \href{https://doi.org/10.1103/PhysRevD.2.1191}{Phys.\ Rev.\ D \textbf{2}, 1191 (1970)}; L.~V.~Keldysh, \href{https://www.jetp.ras.ru/cgi-bin/e/index/e/20/5/p1307?a=list}{Sov.\ Phys.\ JETP \textbf{20}, 1307 (1965)}.

\bibitem{DunneSchubert} G.~V.~Dunne and C.~Schubert, \href{https://doi.org/10.1103/PhysRevD.72.105004}{Phys.\ Rev.\ D \textbf{72}, 105004 (2005)} [\href{https://arxiv.org/abs/hep-th/0507174}{arXiv:hep-th/0507174}].

\bibitem{BLST} C.~Bachas, G.~Lazarides, Q.~Shafi, and G.~Tiktopoulos, \href{https://doi.org/10.1016/0370-2693(91)91597-O}{Phys.\ Lett.\ B \textbf{268}, 401 (1991)} [Erratum: \href{https://doi.org/10.1016/0370-2693(91)90119-B}{Phys.\ Lett.\ B \textbf{271}, 468 (1991)}].

\bibitem{Bachas} C.~Bachas, \href{https://doi.org/10.1016/0550-3213(92)90304-T}{Nucl.\ Phys.\ B \textbf{377}, 622 (1992)}.

\bibitem{RST} V.~A.~Rubakov, D.~T.~Son, and P.~G.~Tinyakov, \href{https://doi.org/10.1016/0370-2693(92)90994-F}{Phys.\ Lett.\ B \textbf{287}, 342 (1992)}.

\bibitem{CMR} E.~Carragher and J.~March-Russell, ``A more effective QCD string at colliders: Decay of excited strings and the worldsheet axion,'' \href{https://arxiv.org/abs/2606.06668}{arXiv:2606.06668}.

\bibitem{NoteJosephson}
Sensitivity of tunneling to a weak drive is familiar from Josephson junctions, where the AC amplitude enters the exponent with an exponentially large coefficient~\cite{IvlevMelnikov}.

\bibitem{NoteRelatedMethods}
Our method has some overlap with~\cite{RST}, which studied transitions from a multiparticle state of fixed energy, and with~\cite{CMR}, where a coherent spectator excitation enters through a boundary condition on a Schwinger--Keldysh contour.

\end{thebibliography}
\end{document}